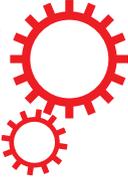

# Atmospheric Beacons of Life from Exoplanets Around G and K Stars

Vladimir S. Airapetian[1,2], Charles H. Jackman[1], Martin Mlynczak[2], William Danchi[1] & Linda Hunt[3,4]

The current explosion in detection and characterization of thousands of extrasolar planets from the Kepler mission, the Hubble Space Telescope, and large ground-based telescopes opens a new era in searches for Earth-analog exoplanets with conditions suitable for sustaining life. As more Earth-sized exoplanets are detected in the near future, we will soon have an opportunity to identify habitale worlds. Which atmospheric biosignature gases from habitable planets can be detected with our current capabilities? The detection of the common biosignatures from nitrogen-oxygen rich terrestrial-type exoplanets including molecular oxygen ($O_2$), ozone ($O_3$), water vapor ($H_2O$), carbon dioxide ($CO_2$), nitrous oxide ($N_2O$), and methane ($CH_4$) requires days of integration time with largest space telescopes, and thus are very challenging for current instruments. In this paper we propose to use the powerful emission from rotational-vibrational bands of nitric oxide, hydroxyl and molecular oxygen as signatures of nitrogen, oxygen, and water rich atmospheres of terrestrial type exoplanets "highlighted" by the magnetic activity from young G and K main-sequence stars. The signals from these fundamental chemical prerequisites of life we call atmospheric "beacons of life" create a unique opportunity to perform direct imaging observations of Earth-sized exoplanets with high signal-to-noise and low spectral resolution with the upcoming NASA missions.

Detection of signatures of life from remote exoplanetary systems is the ultimate goal of astrobiology. The signs of life known as "biosignatures" can potentially be retrieved through the impact of exoplanetary biospheres on atmospheric chemistry. The current methodologies for detecting observable signatures of life in terrestrial-type planets are based on the presence of chemical compounds that are out of chemical equilibrium due to the complex biochemistry of life[1,2]. Signatures of life should meet the criteria of reliability, survivability and detectability and can be identified with remote-sensing detection of the most common molecules in the Earth's troposphere, including molecular oxygen ($O_2$), ozone ($O_3$), water vapor ($H_2O$), carbon dioxide ($CO_2$), nitrous oxide ($N_2O$), and methane ($CH_4$), surface reflectance signatures or polarized scattering in exoplanetary atmospheres[3-6]. The presence of abundant atmospheric molecular oxygen may be a strong indicator of oxygenic photosynthesis, together with methane, the marker of biological decay, and other molecules suggest that the atmosphere is in chemical disequilibrium driven by biological activity[2-4]. However, abiotic production of molecular oxygen can be driven by the photolysis of $CO_2$, $H_2O$ and other oxygen-bearing gases via UV and particle collisions in scenarios of terrestrial-type exoplanets around K, G or M dwarf stars[3,7-9] and could mimic a biologically driven oxygen production. Given this mechanism, the expected $O_2$ production for various scenarios is over 3 orders of magnitude smaller than that observed in current Earth[8]. The situation for very dry (water-depleted) $CO_2$ atmospheres of exoplanets around M dwarf stars is different; in this case $O_2$ can be produced abiotically at the level of Earth-like oxygen abundance[9]. Thus, the absence of water signatures in reflectance and emission spectra can serve as a potential indicator of the abiotic nature of atmospheric $O_2$. In water-rich exoplanetary atmospheres, photolysis of atmospheric water vapor can also produce an accumulation of oxygen due to the thermal escape of hydrogen under some scenarios for exoplanets around active stars[10]. However, the production and accumulation of oxygen could be prevented for terrestrial exoplanets located in the highly compact habitable zones around active M dwarfs. Stellar activity from these host stars can create intense X-ray and Extreme-UV (XUV) emission and magnetized winds, which are the major factors contributing to the non-thermal ionospheric escape of oxygen ions (along with hydrogen)[11].

In general, the abundance of atmospheric molecular oxygen can vary in response of the planetary environments influenced by the fraction of land mass versus ocean mass, volcanic and tectonic activity and the efficiency of atmospheric escape mediated by the activity from the host star[11–13]. Abiotic production of another important biosignature gas, methane, is not an efficient process as compared to its biological origin due to bacteria and biological decay processes, thus making its presence together with molecular oxygen a strong biosignature[14]. However, the current biological production of methane in the Earth's troposphere and stratosphere is mostly due to anthropogenic sources and is ~1.8 ppmv. The major biosignature gases from an Earth-like "living" planet around a Sun-like star are difficult to detect with current telescopes and require long exposures with high spectral, high contrast, and high spatial resolution coronographic instruments on ground-based telescopes such as the Giant-Magellan Telescope (GMT), the Ten Meter Telescope (TMT), the European Extremely Large Telescope E-ELT, and space-based telescopes such as the Habitable Explorere (HabEx), the Large UV Optical and Infrared (LUVOIR), and the Origins Space Telescope (OST)[15,16].

[1]NASA/GSFC, Greenbelt, MD, USA. [2]American University, Washington, DC, USA. [3]NASA/LARC, Hampton, VA, USA. [4]SSAI, Hampton, VA, USA. Correspondence and requests for materials should be addressed to V.S.A. (email: vladimir.airapetian@nasa.gov)

In this paper we propose that strong non-thermal emission from broad molecular bands in the near- and mid-infrared (mid-IR) referred to as "beacons of life" can be emitted by molecular products of nitrogen- and oxygen-rich exoplanetary atmospheres with high (a large fraction of 1 bar) atmospheric pressures. These include stellar activity driven emission from nitric oxide, hydroxyl and oxygen molecules as the strongest signals that may be observable from the habitable worlds.

## Results

**Atmospheric "Beacons of Life" from Earth.** Life as we know it requires, but is not limited to, a nitrogen-rich atmosphere with the presence of abundant free oxygen produced by photosynthesis and the presence of liquid water. Molecular nitrogen ($N_2$) is the dominant gas in the Earth's atmosphere, because of its geochemical stability. The high abundance of atmospheric $N_2$ was critical for the initiation of life on the early Earth, because fixation of nitrogen is required to create linkages of long chained molecules including proteins and base pairs of RNA and DNA[7,17]. Nitric oxide (NO) and hydroxyl (OH) are efficiently produced in the thermosphere as a result of photo-dissociation (*X-ray* and *EUV*) and collisional dissociation (via precipitating electrons) of molecular nitrogen and water vapor (see ***R1-R*** in METHODS). The abundance of NO during quiet times (no geomagnetic storms) varies within a few ppbv and increases by a factor of 100 during strong geomagnetic storms induced by coronal mass ejections from the Sun[18,19]. The abundances of NO, OH and $O_2$ are traced by observations of time-varying emission from molecular bands of NO at 5.3 μm, OH at 1.6 and 2 μm and $O_2$ ($^1\Delta$) at 1.27 μm. NO emission from the 5.3 μm band is created by a large number of vibration-rotation transitions excited by inelastic collisions with atomic and molecular oxygen[19,20] and exothermic reactions. Observations made over the last 15 years with the Sounding of the Atmosphere using Broadband Emission Radiometry (*SABER*) instrument aboard NASA's *TIMED* (Thermosphere Ionosphere Mesosphere Energetics Dynamics) satellite have revealed that the emitting power of NO is strongly correlated with the solar activity cycle traced by the solar radio flux at 10.7 cm, the indicator of solar activity as well as the indices that trace the strength of Coronal Mass Ejection (CME) induced geomagnetic storms. The total power emitted by thermospheric nitric oxide at 5.3 μm and carbon dioxide at 15 μm varies within ~100 GW reaching ~2 TW during strong geomagnetic storms comparable to the one that occurred on Oct 28–31, 2003. The Michelson Interferometer for Passive Atmospheric Sounding (*MIPAS*) observations reveal an increase of NO abundance by a factor of 10 at the height range of 6–68 km during large Oct–Nov 2003 geomagnetic storm reaching a concentration over 300 ppbv at high lattitudes[20]. This caused the enhancement of the NO emission power by a factor of 20–30 in the mid-IR bands as compared to the atmospheric state during the quiet Sun at solar minimum[21]. The drastic enhancement of NO caused the depletion of ozone by up to 60% during the Oct - Nov 2003 storm as observed at high lattitudes at 50–60 km[20].

The radiative power increase during geomagnetic storms is attributed to the enhancement of production of NO molecules and increased kinetic temperature. Infrared NO radiative cooling plays the role of an efficient thermostat by cooling the thermosphere in response to the radiative and non-radiative heating associated with space weather events[22,23]. NO radiative cooling becomes sharply enhanced (more than a factor of 10) due to the increased energy input into the ionosphere-thermosphere (IT) system of polar cap regions during geomagnetic storms. The non-radiative heating is introduced by resistive dissipation of ionospheric currents (Joule heating, JH) in response to the perturbation of the global magnetosphere caused by CME events, ion-neutral collisions and particle precipitation in the polar regions of the planet. If most of the energy entering the IT system is radiated away in the form of NO and $CO_2$ emission, then we can expect that its power should increase proportionally with the input energy flux. The *XUV* flare emission at wavelengths between 1–70 Å is another crucial source of production of low-latitude thermospheric NO driven by photoelectrons formed via photoionization of atmospheric species[23].

The emission in OH bands (1.6 & 2 μm) observed by *SABER* has the total power of ~0.2 TW, which is weakly dependent on the solar cycle (see Fig. 1). Figure 2 presents the flux evolution of $O_2$ ($^1\Delta$) emission flux at 1.27 μm during 2004 to 2008, which shows variations of emission power between zero (at night, increasing substantially in daytime due to $O_3$ photolysis via UV solar radiation) to 230 TW.

**Beacons of Life from Exoplanets Around Active Stars.** Recent Kepler Space Telescope and Hubble Space Telescope data suggest that magnetically active main-sequence stars of K, G and M spectral types generate high XUV fluxes from magnetically driven flares with energies over 10 times greater than that of the current Sun[24,25]. As a result of the energy input, the Earth's thermosphere responds with enhanced radiative cooling from the molecular bands of NO at 5.3 μm, $CO_2$ at 15 μm as well as conductive cooling. Using scaling of NO emission power at 5.3 μm with the F10.7 flux, the proxy for the solar XUV flux, and Ap index, the measure of the geomagnetic activity, we can expect the enhancement of NO emission by over one order of magnitude[20].

Low mass stars including K and M dwarfs are known to remain magnetically active and exhibit high XUV fluxes (over 10 times higher than that of the current Sun) over 2–3 Gyr. For example, XUV fluxes from "mature" 4–6 Gyr-old K and M main-sequence stars are by a factor of 10 greater than that of the Sun at the inner edge of their respective habitable zones[26]. This flux is comparable to the XUV flux from the young 0.7 Gyr-old Sun[7]. This persistent stellar magnetic activity may have a profound effect on the dynamics and chemistry of the early Earth and potential atmospheres of exoplanets around K-M stars[7,11]. Clues to the history of the atmospheric chemistry of the early Earth, specifically its oxygen abundance, come from geological (rock) records suggesting that our planet's atmosphere was weakly oxidizing with less than 0.001% of $O_2$ during the first half of our planet history including the time when life started on Earth. Only ~2.4 Gyr ago our atmosphere acquired appreciable concentration of free oxygen. This puzzling increase of $O_2$ is known as Great Oxidation Event (GOE)[27]. The GOE time scale is comparable or shorter than the magnetic evolution time scale for low mass cool stars. Can stellar activity control the onset of the GOE on terrestrial-type exoplanets? Oxygenated atmospheres of terrestrial type exoplanets can be created by several factors including atmospheric escape, surface gravity, and the magnetic activity of its host star. One of the factors causing the accumulation of atmospheric oxygen in the troposphere at ~2.4 Gyr ago could be XUV-driven photolysis of the carbon dioxide the low pressure (~0.25 bar) early Earth atmosphere depleted due to XUV driven escape of oxygen and nitrogen ions[11,28]. This may suggest that the active low mass stars with greater XUV fluxes can induce comparable atmospheric erosion at shorter time scale causing $O_2$ accumulation earlier in the history of a terrestrial type exoplanet. Thus, in many scenarios terrestrial type exoplanets with $N_2$ – rich oxygenated atmospheres can be exposed to high energy fluxes from their host stars. XUV emission and energetic particles can be efficient in photodissociation of the atmospheric molecular nitrogen, carbon dioxide and water vapor resulted in production of nitric oxide and hydroxyl molecules[7,20,21].

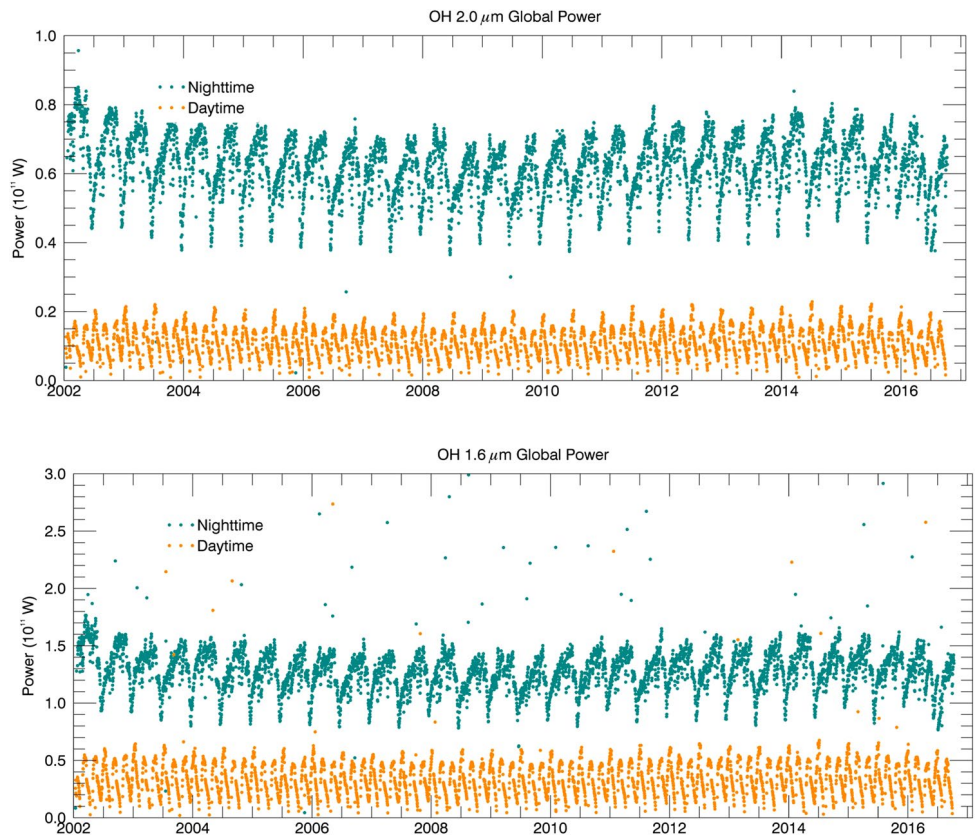

**Figure 1.** Total emission power (in units of 0.1 TW): upper panel: OH (9→7 + 8→6) at 2 μm; lower panel: OH (5→3 + 4→2) power at 1.6 μm.

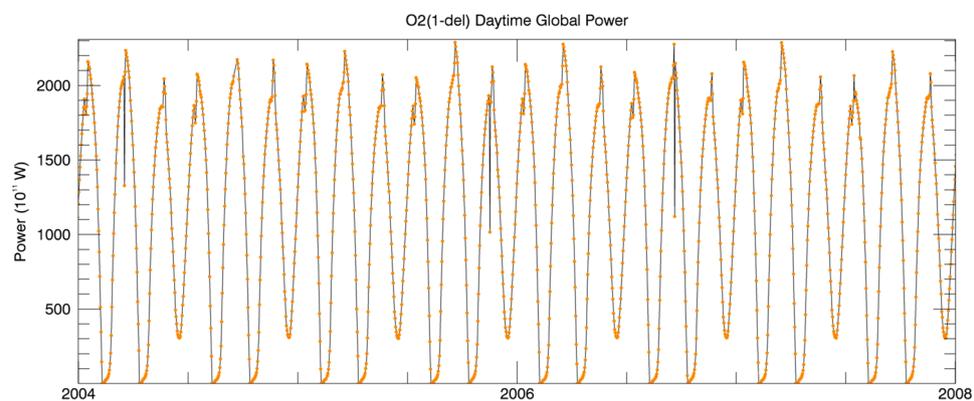

**Figure 2.** The daily averages of the total emitting power (in units of 0.1 TW) by the Earth as a dot from $O_2(^1\Delta)$ at 1.27 μm.

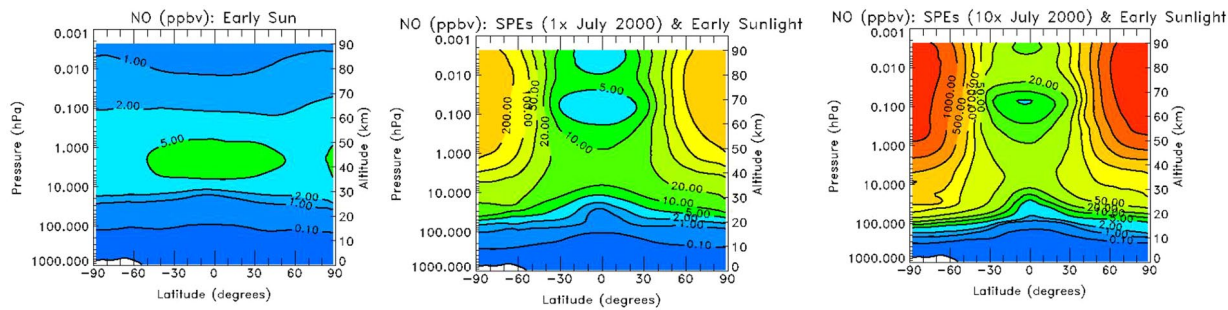

**Figure 3.** 2D map of steady state NO mixing ratio (in ppbv): Left: Model B output, Middle: Model C output and Right: Model D output; Contour level: 0.1, 1, 2, 5, 10, 20, 50, 100, 200, 500, 1000, & 2000 ppbv.

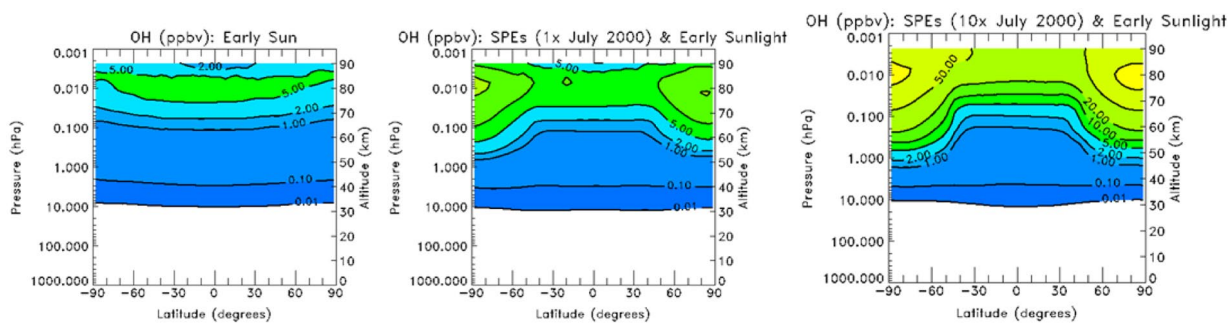

**Figure 4.** 2D map of steady state OH mixing ratio (in ppbv): Model B (left panel), Model C (middle panel) and Model D (right panel); Contour level: 0.1, 1, 2, 5, 10, 20, 50, 100 ppbv.

How would space weather from active stars in the form of XUV radiation and SEP particle fluxes affect the amount of NO and OH emission in the mid-IR bands that would be expected from oxygen-rich terrestrial-type exoplanets around active G, K and M stars? To investigate the atmospheric chemistry changes owing to CME driven solar energetic particles (SEPs), we expose an Earth-like oxygen-rich atmposphere to the energy fluxes expected from active G, K and M main-sequence stars. We computed a grid of 5 atmospheric chemistry models of an Earth twin to numerically study the chemical evolution of nitrogen oxides, $NO_x$ (N, NO, $NO_2$) $HO_x$ (H, OH and $HO_2$) constituents and ozone in response to various fluxes of energetic protons during SEPs in the stratosphere and themosphere. For this purpose, we used the two-dimensional photochemical transport Goddard Space Flight Center (*GSFC*) atmospheric model that calculates the constituents in the upper stratosphere (35 km) extending to the upper mesosphere (95 km) (see METHODS). Our study includes the base model (base Model A, no SEP and the current XUV flux, $F_0$), Model B (no SEP and XUV flux at $10F_0$), Model C (July 2000 SEP event and XUV flux at $10F_0$), Model D (10 × July 2000 SEP event and XUV flux at $10F_0$), Model E (20 × July 2000 SEP event and XUV flux at $10F_0$). We used the maximum daily average particle input for the SEP event, which occurred on July 15, 2000, that was associated with the fast (1800 km/s) halo CME and the Bastille Day × 5.7 solar flare[29].

Our simulations suggest that at the atmospheric height range between 0 to 95 km, the enhanced XUV flux from the Sun is not efficient in the production of odd nitrogen, and thus only modest changes in NO mixing ratio are observed. Model B (no SEP and the active Sun at $10F_0$) shows NO values of typically 1 ppbv at 80–90 km (the left panel of Fig. 3), which are similar to those computed for base Model A. The steady state energy flux in the form of energetic protons from 1–300 MeV with the energy spectrum and intensity characteristic of the July 2000 SEP event (Model C, see also Methods) produces a mixing ratio for NO that is enhanced by a factor of 200 (or 200 ppbv at the reference height of 85 km) in the polar cap regions (the middle panel of Fig. 3) with respect to the quiet conditions described by Model B. Model D suggests that NO net production increases approximately linearly in response to the enhancement of the intensity of the proton flux by a factor of 10 with respect to Model C, and reaches 2 ppmv (the right panel of Fig. 3). These fluxes are consistent with the results of 1D atmospheric models applied for SEP driven atmospheric chemistry of exoplanets around M dwarfs[30]. Such large concentrations of NO from an oxygen-rich atmosphere are by a factor of 10–30 greater than than modeled in the abiotic atmosphere of the early Earth[7]. Thus, the false positives do not appear to be important production of NO as an atmospheric "beacon" of life. Model E results show that this trend continues as we increase the proton flux by another factor of 2. Thus, we conclude that the net production of NO is scaled approximately linearly with the incident proton flux. This could be understood, because 70 to 80% of the energy of incoming protons goes into creation of ionization pair production. Neutral and ion chemistry resulting from ion pair production creates about 1.25 N atoms and about 2 $HO_x$ molecules for each incident proton, and the N atoms produced are in the ground state $N(^4S)$ (45% per ion pair) or the excited state $N(^2D)$ (55% per ion pair)[31].

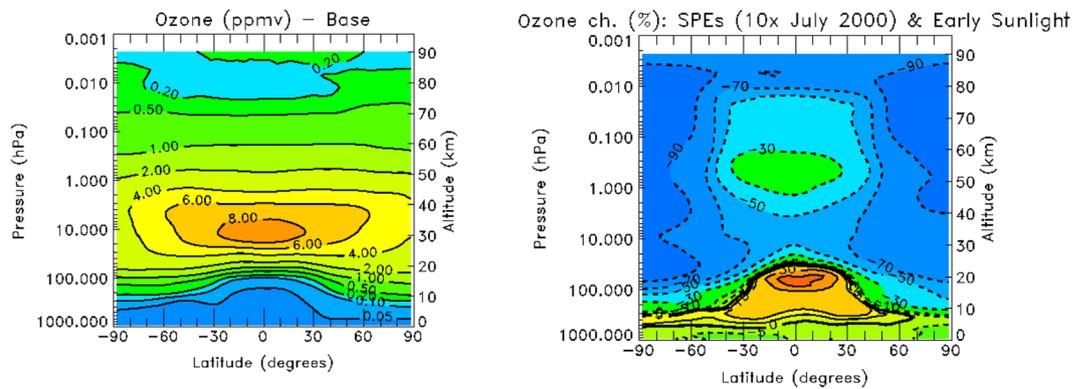

**Figure 5.** The 2D map of atmospheric ozone destruction driven by XUV flux from an active star and a 10xJuly 2000 SEP event (right panel, Model D) with respect to the base model (left panel; Model A).

The increase in the NO concentration by 2 to 3 orders of magnitude is driven by the corresponding flux of precipitating protons produced by CME initiated shocks and associated secondary electrons. Each SEP associated CME event exerts a dynamic pressure on the planetary magnetosphere, which perturbs the geomagnetic field and induces geomagnetic currents in the polar regions. The resistive dissipation of these currents heats the ionosphere. The height integrated Joule heating (JH) rate introduced by such currents varies from a few mW/$m^2$ (for a regular CME event with the energy of $10^{31}$ ergs) to up to 1 W/$m^2$ for a super Carrington event with the energy of $2 \times 10^{33}$ ergs[7,32]. The JH rate increased by 2 orders of magnitude should be balanced by the radiative and conductive cooling rates. The radiative cooling rates are mostly controlled by greater emission flux from NO molecules due to enhanced production of NO due to SEP mediated dissociation. The generation of large amounts of NO has been shown in Earth's atmosphere to lead to "overcooling" events[33]. That is, so much NO is generated that the resultant infrared cooling leaves the atmosphere colder than it was before the geomagnetic event. NO emission at 5.3 μm is driven by collisional impact with atomic oxygen, and therefore, is dependent on thermospheric abundance of O. The models show no significant change in production of atomic oxygen due to XUV flare emission from the active star (at $10F_0$) at altitudes below 70 km (Model B). The proton events contribute to the modest (20–40%) destruction of O in the thermosphere. However, the NO emission flux at 5.3 μm depends on the temperature of neutral oxygen, $T_n$, as ~exp$(-2700/T_n)$, because the increase in neutral temperature increases the rate of collisional excitation of NO, while reduction of O plays a secondary role in the production of NO emission flux[34]. The neutral temperature increase is caused by the enhaced Joule heating deposited in the thermosphere during geomagnetic storms, and thus mostly contributes to the rise in NO emission power.

The $O_2$ signal is proportional to J $[O_3]$, where J is the ozone photolysis rate, and the symbol $[O_3]$ represents the abundance of $O_3$. The ozone abundance also depends on more $O_2$ photolysis (see ***R11*** of METHODS), it is inversely dependent on temperature, and also on the H, NO, $NO_2$, and other chemical abundances (see ***R13*** of METHODS). Figure 5 presents the 2D map of atmospheric ozone destruction driven by XUV flux from an active star and a 10xJuly 2000 SEP event (right panel, Model D) with respect to the base model (left panel; Model A). The Models C and D show that ozone is destroyed up to 70% in the mid latitudes and up to 90% at latitudes over 60°. This is caused by both ozone photolysisand collisional dissociation with energetic particles and enhanced production of NO. The NO radiative cooling is sentitive to the temperature of neutral oxygen as discussed above, and thus, the rapid cooling cuts off the infrared emission – this is a manifestation of the "thermostat" effect discussed above. More detailed modeling of the response of Earth-like planets is required to understand whether NO emission would be prolonged after storms or rather would be short-lived due to overcooling.

The same linear scaling is observed for the increase of steady state resultant mixing ratios of production of $HO_x$ (H, OH and $HO_2$) due to dissociation of water vapor via the incoming flux of SEP protons. Figure 4 shows the 2D maps of the OH mixing ratios for the models B, C and D (as in Fig. 3). One can see that the mixing ratio for Model C increases to 100 ppbv at the reference height. As discussed in reference[35], most of the OH produced is the vibrationally excited hydroxyl molecule; thus the expected power emitted by the 1.6 μm and 2.0 μm bands should increase accordingly by the same factor if a sufficient amount of ozone is available for its production.

We point out that understanding the response of the upper atmosphere of a planet to strong CMEs is still a frontier of research even for the Earth. The current Decadal Survey [http://nap.edu/13060] for Solar and Space Physics identifies this topic as a Key Science Goal for Earth's thermosphere and ionosphere.

### Discussion

Our results imply that the total NO emitting flux at 5.3 μm from the planet around an active K to G star (Models C and D) with an N-$O_2$ rich atmosphere is expected to be ~$10^{20}$–$10^{22}$ erg/s. This is consistent with the estimate of the NO power at 5.3 μm during higher stellar activity expected from younger G and K type stars obtained from the linear scaling between X-ray flux (0.1–0.8 nm) GOES measurements with F10.7 flux obtained in[18] extended to F10.7 = 2000. The contribution of the false positive of emitted flux is less than the NO production rate by over 3-4 orders of magnitude, because the emission of NO is mediated by the collsions of atomic O that is strongly suppressed in the abiotic atmosphere[8]. We should note that our model does not account for the thermospheric temperature caused by the energy inputs due to XUV emission and JH, and thus, represents a low bound of expected NO emission fluxes at 5.3 μm from nitrogen- and oxygen rich atmospheres of exoplanets around active stars. The emission power from OH bands $-2 \times 10^{18}$–$10^{20}$ ergs/s and $O_2$ at 1.27 μm at least at the level of $2 \times 10^{21}$ ergs/s.

For a planetary system located at 10 pc away from Earth, this implies the total flux at the Earth integrated over the NO band with the bandwidth of ~5600 Å at $10^{-20}$–$10^{-18}$ ergs/cm$^2$/s, the OH bands with the bandwidth of ~2000 Å at $2 \times 10^{-22}$–$10^{-20}$ erg/cm$^2$/s and $O_2$ band over the spectral width of 431 Å at $\sim 2 \times 10^{-19}$ ergs/cm$^2$/s. This suggests that low spectral resolution $\lambda/\Delta\lambda = 10$–$40$ would be required to detect emission from these bands. Low spectral resolution ($\lambda/\Delta\lambda \sim 20$) observations at a signal-to-noise of 10 will require only 1.5 hours of exposure time with the JWST/MIRI spectrograph at ~5 μm. Thus, if future space telescopes have apertures larger than JWST by a factor of 2, the exposure time could be reduced by a factor of 4 and the planets with the 10 times lower NO flux can be detected in just 4 hours. Because the emission flux from NO, OH and $O_2$ ($^1\Delta$) bands are dependent on stellar activity levels driven by geomagnetic storms & XUV flux, it is expected to vary by a factor of 10 to 100 depending on the level of stellar activity on a time scale of storms, or ~2–3 days. This variability time scale should be greater than the expected exposure time. The estimated number of prime Earth twin targets around G-K stars in the neighborhood of 10 pc for the detection of the beacons of life is at least ~20 planets[36,37].

## Conclusion

Based on our analysis of *TIMED/SABER* data collected from observations of Earth's thermospheric emission in the mid-IR bands and 2D models of atmospheric chemistry of terrestrial type planets around active stars, we conclude that the molecular-band emission from NO at 5.3 μm, OH at 1.66 and 2 μm and $O_2$ at 1.27 μm is strongly correlated with the level of stellar activity in the form of stellar energetic particle events. The high emission fluxes from these molecules in the mid-IR bands would signify a nitrogen-rich, oxygen rich and water rich atmosphere with the atmospheric pressure (significant fraction of 1 bar), which represent the fundamental requirements for habitability referred to as atmospheric "beacons of life". Our models suggest that ozone in the atmospheres of habitable exoplanets with high NO and OH fluxes should significantly depleted at all lattitudes via reactions with $NO_x$ and $OH_x$ species (see Fig. 5).

Although the raw signal itself from NO is observable as noted above, however, its size is small compared to the thermal emission from the exoplanet, which is $\sim 10^2$ larger, and much smaller than the emission from the host star, which can be $\sim 10^6$ greater for a star similar to the Sun. Thus, a possible observational strategy available in the near term, using *JWST*, would be to perform eclipse spectroscopy, where thermal emission from the exoplanet is observed[38,39]. Potential targets could be found by *TESS* and then observed with *JWST*. Ideally the host star would be known to be active, and, if possible, observed close in time by optical or UV observatories to correlate the eclipse observations with recent activity. The proposed time-varying "atmospheric beacons of life" can be observed using direct imaging techniques in the mid IR bands via recently proposed "Exo-Life Beacon Space Telescope", an extended version of *FKSI* space interferometer[39,40].

## Methods

**2-D GSFC Atmospheric Model.** We employed the Goddard Space Flight Center (GSFC) two-dimensional (2-D) atmospheric model to study the impact of solar protons on the Earth atmospheric chemistry[41,42]. The vertical range of the model is equally spaced in log pressure and extends from the ground to approximately 92 km (0.0024 hPa) with a 1 km grid spacing and 4 degree latitude grid spacing. The transport is computed off-line and is derived using the daily average global winds and temperatures from the NASA Modern Era Retrospective-analysis for Research and Applications (MERRA) meteorological analysis (see the website: http://gmao.gsfc.nasa.gov/research/merra/) for 1979–2012. Thirty-day running averages of the residual circulation, eddy diffusion, zonal mean wind, and zonal mean temperature are computed using the methodology described in[43] and are used as input into the GSFC 2-D model.

The ground boundary conditions in the GSFC 2-D model for the source gases are taken from WMO (2011) for year 2000. The model uses a chemical solver described in[43–45]. The photochemical gas and heterogeneous reaction rates and photolysis cross sections have been updated to the Jet Propulsion Laboratory recommendations from year 2010 for these computations[46]. We use the solar proton flux (energies 1 to 300 MeV) for the July 14-16, 2000 SEP event provided by the National Oceanic and Atmospheric Administration (NOAA) Space Weather Prediction Center (SWPC) for the NOAA Geostationary Operational Environmental Satellites (GOES) (see http://www.swpc.noaa.gov/ftpmenu/lists/ particle.html). The GOES 13 data are considered to be the most reliable of the current GOES datasets for the proton fluxes depositing energy into polar latitudes and were used as the source of protons in several energy intervals for the July 2000 SEP event (see also at http://www-istp.gsfc.nasa.gov/istp/events/2000july14/20000716_proton.gif). Peak ionization rates above 10000 cm$^{-3}$ s$^{-1}$ and the energy deposition rate of $5 \times 10^{-7}$ ergs/cm$^3$/s were reached on July 15, 2000.

The proton flux data were used to compute the ion pair production profiles employing the energy deposition methodology discussed in[46], where the creation of one ion pair was assumed to require 35 eV. The SEP-produced hourly average ionization rates and the energy deposition rates in the polar cap regions (>60 geomagnetic latitude) are given in Figs 6 and 7, respectively, for July 14–16 of 2000 SEP event. The GOES proton flux during the July 14–16 of 2000 event as well as the impact on the atmosphere is discussed in[47]. We use the daily average ionization rates for July 15, 2000 (the maximum energy deposition day of this SEP event) in the computations shown in the paper.

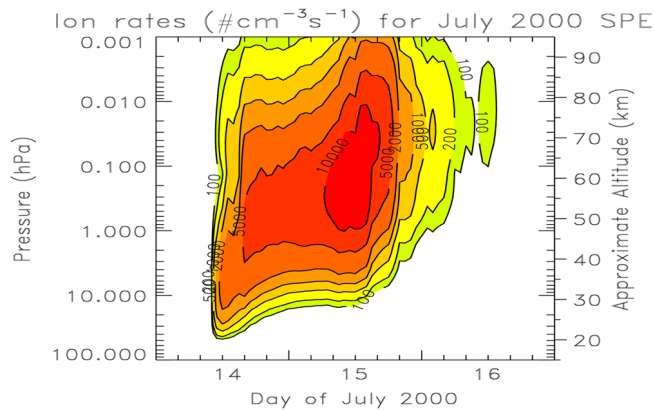

**Figure 6.** Ionization rates (in units of # cm$^{-3}$ s$^{-1}$) for July 2000 SPE event.

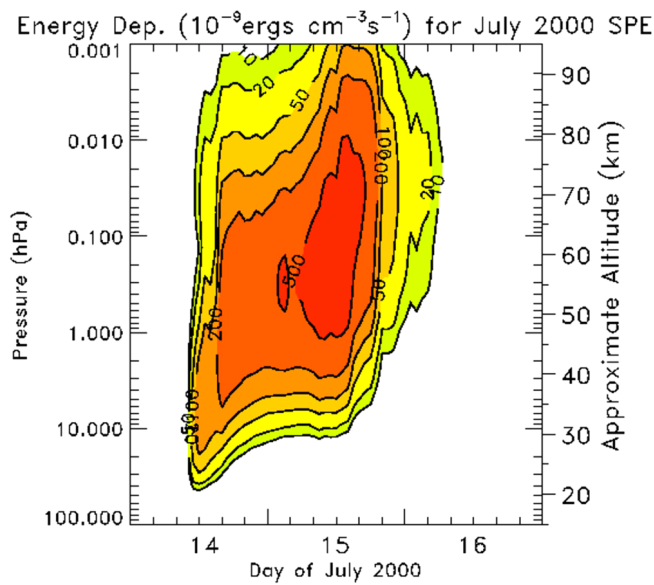

**Figure 7.** Energy deposition (in units of 10$^{-9}$ ergs cm$^{-3}$ s$^{-1}$) for July 2000 SPE.

This model referred to as Model A serves as the reference model for calculation of atmospheric chemistry for Model B with 10x enhanced XUV flux and Models C, D, and E for 1x, 10x, and 20x of the particles fluxes, respectively.

**Production and Destruction of NO and OH.** Production and destruction of NO and OH species are driven by the following reactions

$$N_2 + e^- \rightarrow 2N(^4S) + e^- \quad (R1)$$

$$N_2 + h\nu \rightarrow N(^4S) + N(^2D) \quad (R2)$$

Other possibilities from this interaction are

$$N_2 + e^- \rightarrow N(^4S) + N(^2D) + e^- \quad (R3)$$

as well as other products, see[9,10]. The dominant sources of NO production are

$$N(^4S) + O_2 \rightarrow NO + O(^3P) \quad (R4)$$

$$N(^2D) + O_2 \rightarrow NO + O(^3P), \quad (R5)$$

and destruction through

$$N(^4S) + NO \rightarrow N_2 + O(^3P). \tag{R6}$$

(R4) and (R5) are the major sources of production of NO that becomes vibrationally excited by impacts with atomic oxygen in the ionosphere-thermosphere (IT), and thus are temperature dependent. The hydroxyl production is controlled by photolysis of water and collisional dissociation

$$H_2O + Ion^+ \rightarrow H + OH + Ion^+ \tag{R7}$$

$$H + O_3 \rightarrow OH + O_2. \tag{R8}$$

### Destruction of Thermospheric Ozone.
Ozone production comes from the following reactions

$$O_2 + h\nu(<242\ nm) \rightarrow O + O \tag{R9}$$

followed by

$$O + O_2 + M \rightarrow O_3 + M \tag{R10}$$

Ozone destruction is mediated via the following reactions

$$O_3 + h\nu(<255\ nm) \rightarrow O_2 + O \tag{R11}$$

followed by

$$O + O_3 \rightarrow O_2 + O_2 \tag{R12}$$

Catalytic destruction of ozone is driven by the following reactions

$$O_3 + X \rightarrow XO + O_2 (X=O, NO, OH, Br\ or\ Cl) \tag{R13}$$

$$O + XO \rightarrow X + O_2 \tag{R14}$$

$$Net: O + O_3 \rightarrow O_2 + O_2 \tag{R15}$$

X is not destroyed but continuously participates in the cycle.

### Data availability statement.
Observational data on OH and $O_2$ emission fluxes are available from TIMED/SABER mission at https://gcmd.nasa.gov/records/GCMD_TIMED_SABER.html. The 2D GSFC atmospheric model data are available at NASA Goddard Space Flight Center. The results of simulations are available upon request from the corresponding author.

### Acknowledgements

The authors would like to thank the referees for constructive suggestions that improved and clarified the major findings of this paper. VA's work was supported by internal funding from NASA GSFC's Sellers Exoplanetary Environments Collaboration (SEEC) and by NASA's Exobiology program, Grant No. 80NSSC17K0463. VA performed the part of this work while staying at ELSI/Tokyo Institute of Technology, Japan.

### Author Contributions

V.S.A. conceived and designed the numerical experiments, analysed the data, contributed materials and wrote the manuscript. C.J. contributed to the development and execution of codes and data analysis, M.M. and L.H. contributed to the data analysis, W.D. contributed to development of the manuscript and to data analysis.

### Additional Information

**Competing Interests:** The authors declare that they have no competing interests.

**Publisher's note:** Springer Nature remains neutral with regard to jurisdictional claims in published maps and institutional affiliations.